\documentclass[10pt,conference]{IEEEtran}
\IEEEoverridecommandlockouts
\usepackage{amsmath,amssymb,amsfonts}
\usepackage{algorithmic}
\usepackage[ruled]{algorithm2e}
\usepackage{array}
\usepackage{textcomp}
\usepackage{stfloats}
\usepackage{url}
\usepackage{color}
\usepackage{verbatim}
\usepackage{graphicx}
\usepackage{tabularx}
\usepackage{multirow}
\usepackage{array}
\usepackage{booktabs}
\usepackage{graphicx}
\usepackage{parskip}

\newtheorem{definition}{Definition}
\usepackage[numbers]{natbib}
\usepackage{bm}
\usepackage{mathtools}

\usepackage{subfigure}

\usepackage{bibspacing}
\graphicspath{ {./fig/} }
\usepackage{enumitem}

\usepackage{flushend}
\def\BibTeX{{\rm B\kern-.05em{\sc i\kern-.025em b}\kern-.08em
    T\kern-.1667em\lower.7ex\hbox{E}\kern-.125emX}}
\begin{document}

\title{Composite and Staged Trust Evaluation for Multi-Hop Collaborator Selection} 

\author{\IEEEauthorblockN{Botao Zhu and Xianbin Wang}
\IEEEauthorblockA{Dept. of Electrical and Computer Engineering, Western University,
London, Ontario N6A 3K7 CANADA \\}}




\maketitle

\begin{abstract}

   Multi-hop collaboration offers new perspectives for enhancing task execution efficiency by increasing available distributed collaborators for resource sharing. Consequently, selecting trustworthy collaborators becomes critical for realizing effective multi-hop collaboration. However, evaluating device trust requires the consideration of multiple factors, including relatively stable factors, such as historical interaction data, and dynamic factors, such as varying resources and network conditions. This differentiation makes it challenging to achieve the accurate evaluation of composite trust factors using one identical evaluation approach. To address this challenge, this paper proposes a composite and staged trust evaluation (CSTE) mechanism, where stable and dynamic factors are separately evaluated at different stages and then integrated for a final trust decision. First, a device interaction graph is constructed from stable historical interaction data to represent direct trust relationships between devices. A graph neural network framework is then used to propagate and aggregate these trust relationships to produce the historical trustworthiness of devices. In addition, a task-specific trust evaluation method is developed to assess the dynamic resources of devices based on task requirements, which generates the task-specific resource trustworthiness of devices. After these evaluations, CSTE integrates their results to identify devices within the network topology that satisfy the minimum trust thresholds of tasks. These identified devices then establish a trusted topology. Finally, within this trusted topology, an A* search algorithm is employed to construct a multi-hop collaboration path that satisfies the task requirements. Experimental results demonstrate that CSTE outperforms the comparison algorithms in identifying paths with the highest average trust values.

\end{abstract}

\begin{IEEEkeywords}
   GNN, staged evaluation, integrated decision, multi-hop, path planning 
\end{IEEEkeywords}

\section{Introduction}

As interconnected systems and applications grow more complex, individual devices often struggle to manage computationally heavy tasks due to their limited processing power and energy resources. To address this limitation, the concept of distributed resource scheduling has gained attention, allowing tasks to be offloaded to more powerful devices through multi-hop relays~\cite{X. Wang}. For instance, in industrial automation, collaborative robots utilize multi-hop communication to coordinate complex assembly lines, while in smart cities, autonomous vehicles leverage vehicle-to-vehicle and vehicle-to-infrastructure multi-hop networks for cooperative perception and task execution. A critical prerequisite for successful task completion in such scenarios is the selection of reliable relay and computing devices. However, the heterogeneous and intricate nature of device resources—encompassing varying CPU capacities, energy levels, network conditions, and historical performance—renders traditional methodologies, such as rule-based assessments, inadequate for effectively evaluating device reliability~\cite{bzhu3},\cite{Z. Xiao}.

Trust has emerged as a key metric for evaluating the reliability of collaborators in multi-hop collaboration systems, where it is defined as a collaborator's ability to fulfill the task requirements specified by the task owner. Some studies assess device trust by analyzing multi-dimensional historical data, such as task completion rates, packet loss rates, and computation delays, and then applying predefined rules or machine learning models to calculate trust~\cite{B. Zhu1},\cite{M. M. E. A. Mahmoud}. Other studies focus on analyzing historical interaction patterns between devices—such as shared interests and behavioral similarities—to quantify trust relationships~\cite{B. Zhu2}. However, these approaches often fail to achieve accurate trust evaluations in complex and distributed systems. First, they focus on evaluating trust based on historical data, which may not yield accurate results under dynamic conditions. In fact, to achieve accurate trust evaluation, factors such as available device resources should also be considered as trust-determining elements. Second, they typically employ a single evaluation mechanism to assess all trust-related factors, overlooking the varying nature of some factors. For instance, historical interaction data tends to be stable, whereas computational resources are highly dynamic. Thus, there is an urgent need for a new mechanism capable of handling both stable and dynamic factors to achieve accurate and comprehensive trust evaluation.

In this research, we propose a composite and staged trust evaluation (CSTE) mechanism, which separately evaluates stable and dynamic factors and integrates them to make the final trust decision. To begin, stable historical interaction data is used to create a device interaction graph, which models direct trust relationships. This graph is then processed by a graph neural network (GNN) framework to produce the historical trustworthiness of devices through propagation and aggregation. 
Furthermore, a task-specific trust evaluation method is developed, assessing dynamic device resources based on task requirements and resulting in task-specific resource trustworthiness for devices. Upon obtaining the results of these evaluations, CSTE integrates them to identify devices within the network topology that meet the minimum trust thresholds. These identified devices then constitute a trusted topology. Finally, an A* search algorithm is employed within this trusted topology to build a multi-hop collaboration path fulfilling the task requirements.

\section{System Model and Problem Formulation}

\subsection{System Overview}

A collaborative system is considered, comprising a set of terminal devices $A = \{a_1, \dots, a_I\}$ and edge computing (EC) devices $B = \{b_1, \dots, b_M\}$. Terminal devices, such as mobile phones and robots, can act as task initiators that generate tasks or as task forwarding (TF) devices that assist in relaying tasks. EC devices, equipped with computational capabilities, provide computing services to terminal devices. The overall network topology is modeled as $G^{\text{top}} = ((A, B), E^{\text{top}})$, where $E^{\text{top}}$ denotes the set of communication links among devices. Each link is represented by $e_{a_i,a_j}$ or $e_{a_i,b_m}$, corresponding to a connection between two terminal devices or between a terminal device and an EC device, respectively. Device $a_i$ is assumed to be a task initiator, generating a task $C$, which is parameterized as ($c^{\text{des}}$, $c^{\text{size}}$, $c^{\text{TF}}$, $c^{\text{EC}}$), where $c^{\text{des}}$ represents the processing density (cycles/bit), $c^{\text{size}}$ denotes the number of data bits, $c^{\text{TF}}$ is the minimum trust threshold for TF devices, $c^{\text{EC}}$ represents the minimum trust threshold for EC devices. Due to geographical constraints,  task $C$ must be relayed through multiple trusted terminal devices before reaching a trusted EC device. Additionally, the system deploys monitoring devices to collect data from devices involved in collaboration.

\subsection{Trust Model}

To ensure effective and reliable completion of task $C$, all selected collaborative devices on the multi-hop path must satisfy the trust thresholds specified by $C$. Therefore, it is necessary to evaluate the trustworthiness of both terminal and EC devices. 
We respectively define the trust evaluation models for terminal devices and edge computing devices as follows. 

\begin{definition}[Task forwarding trust]
    The task forwarding trust that task initiator $a_i$ places in a terminal device $a_j$ is defined as $a_i$'s expectation of $a_j$'s ability to successfully forward task $C$, which is calculated as
    \vspace{-0.08 in}
    \begin{equation}
        \label{final_a_i_a_j}
        T_{a_i,a_j} = T_{a_i,a_j}^{\text{his}} (D) T^{\text{res}}_{a_i,a_j}(C), 
    \end{equation}
    where $T_{a_i,a_j}^{\text{his}}$ represents the historical trustworthiness of $a_j$ derived from the collected historical interaction data $D$, and $T^{\text{res}}_{a_i,a_j}$ reflects the resource trustworthiness of $a_j$ specific to task $C$. 
\end{definition} 
\begin{definition}[Task computing trust]
The task computing trust that task initiator $a_i$ places in an EC device $b_m$ is defined as $a_i$'s expectation of $b_m$'s ability to execute task $C$, which is calculated as
   \vspace{-0.1 in}
    \begin{equation}
        \label{final_a_i_b_m}
        T_{a_i, b_m} = T_{a_i, b_m}^{\text{his}}(D) T^{\text{res}}_{a_i, b_m} (C),  
    \end{equation}
where $T_{a_i,b_m}^{\text{his}}$ denotes the historical trustworthiness of $b_m$ based on historical interaction data $D$, while $T^{\text{res}}_{a_i,b_m}$ is the resource trustworthiness of $b_m$ for task $C$. 
\end{definition}

According to Definitions 1 and 2, evaluating the trustworthiness of a device requires analyzing not only its historical interaction data but also whether its available resources meet the requirements of task $C$. We assume that interaction $d_{a_i,a_j} \in D$ represents an instance where $a_j$ assists in forwarding a task from $a_i$, recording the relevant performance of $a_j$ during this interaction. An interaction from $a_i$ to $b_m$, in which $b_m$ assists in computing a task generated by $a_i$, is denoted as $d_{a_i,b_m} \in D$, which captures the relevant performance metrics of $b_m$ during task execution. 
 
 \vspace{-0.03 in}
 \subsection{Problem Formulation}
 According to the network topology $G^{\text{top}}$, all paths from $a_i$ to $M$ EC devices are assumed to be represented as the set ${\Phi}$. One of the paths, $\pi$, is assumed to contain $(K-1)$ TF devices and one EC device $b_m$. The sum of trust values of these $K$ devices is calculated as $ T = \sum_{}^{K-1} T_{a_i,a_j} + T_{a_i,b_m}$. Considering the reliability of both task transmission and task computation processes, this study aims to plan a multi-hop path from task initiator to an EC device such that the average trust value of all devices on the path is maximized. The optimization problem is formulated as
\vspace{-0.05 in}
\begin{align}
     \label{problem}
     &\max_{{\Phi}, {A}, {B}} \quad \frac{T}{K},   \\
    \mathrm{s.t.} \quad
    &  T_{a_i, a_j} \ge c^{\text{TF}}, a_j \in \pi, a_j \in A, \label{sub-1}\\
    &  T_{a_i, b_m} \ge c^{\text{EC}}, b_m \in \pi, b_m \in B, \label{sub-2}\\
    & \pi \in \Phi. \label{sub-3}
\end{align}
 Constraint (\ref{sub-1}) states that the trustworthiness of the selected TF devices should meet the minimum trust threshold of $C$ for TF devices. Constraint (\ref{sub-2}) specifies that the trustworthiness of the selected EC device must meet the minimum trust threshold of $C$ for EC devices.

\section{Independent Trust Evaluation and Integrated
Decision for Multi-Hop Collaborator Selection}

To address Problem (\ref{problem}), this study proposes the CSTE mechanism. It first uses the GNN-enabled trust evaluation framework to compute the historical trustworthiness of devices based on stable historical interaction data. Then, a task-specific trust evaluation approach is applied to assess the dynamic resource trustworthiness of devices. The results of these two evaluations are then integrated to identify trustworthy collaborators. Finally, the A* search algorithm is employed to efficiently construct a multi-hop collaboration path with the highest average trust.

\begin{figure*}[t!]
\centering
\includegraphics[scale=1.01]{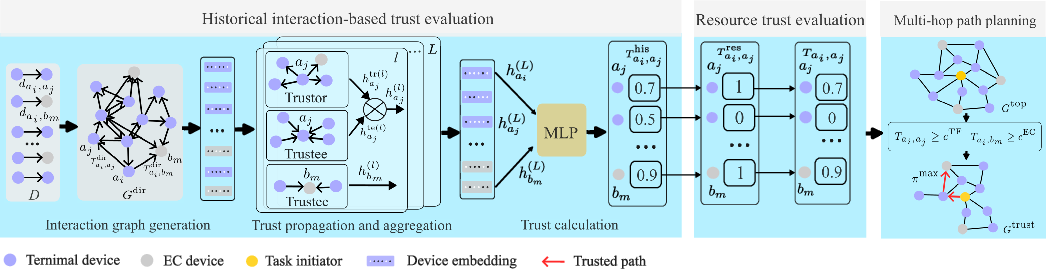}
\caption{The proposed CSTE mechanism plans a multi-hop collaboration path with the maximum average trust value.}
\label{gnn}
\end{figure*}

\subsection{Historical Interaction-Based Trust Evaluation}

To obtain the historical trustworthiness of devices, we implement a GNN-enabled evaluation framework that consists of three steps: interaction graph generation, propagation and aggregation of trust information, and trust calculation.

\textit{1) Interaction Graph Generation}: According to the collected historical interaction data $D$ between devices, their direct trust relationships can be computed. To comprehensively assess the capabilities of TF devices, we use the packet loss rate (PLR) to evaluate the quality of communication links and the task forwarding success rate (TFSR) to measure their forwarding capability. Hence, an interaction $d_{a_i,a_j}$ encompasses $a_j$'s PLR and TFSR. Accordingly, an edge $e_{a_i, a_j}$ can be established from $a_i$ to $a_j$. The weight of $e_{a_i, a_j}$ indicates the direct trust $T_{a_i, a_j}^{\text{dir}}$ that $a_i$ places in $a_j$, computed based on all interactions from $a_i$ to $a_j$ as follows
\begin{align}
    \label{a_i_dir_a_j}
    T_{a_i, a_j}^{\text{dir}} = \alpha_1 T_{a_i, a_j}^{\text{PLR}} +  \alpha_2 T_{a_i, a_j}^{\text{TFSR}}, 
\end{align}
where $\alpha_1$ and $\alpha_2$ are weight coefficients, $0 \leq \alpha_1, \alpha_2 \leq 1$, $ \alpha_1 + \alpha_2 = 1$. $T_{a_i, a_j}^{\text{PLR}}$ and $T_{a_i, a_j}^{\text{TFSR}}$ measure the communication link from $a_i$ to $a_j$ and the packet forwarding ability of $a_j$, respectively. These metrics are calculated as 
\vspace{-0.08 in}
    \begin{align}
    T_{a_i, a_j}^{\text{PLR}} &=  \frac{1}{N_{a_i, a_j}}\sum_{n=1}^{N_{a_i, a_j}}\left(1 - \frac{P^{\text{lost}}_n}{P^{\text{tot}}_n}\right), \\
    T_{a_i, a_j}^{\text{TFSR}} &= \frac{1}{N_{a_i, a_j}}\sum_{n=1}^{N_{a_i, a_j}} \frac{P_n^{\text{tra}}}{P_n^{\text{rec}}},
    \end{align}
where $N_{a_i, a_j}$ is the total number of tasks received by $a_j$ from $a_i$. $P^{\text{tot}}_n$ denotes the total number of packets sent from $a_i$ to $a_j$ in the $n$-th task, and $P^{\text{lost}}_{n}$ indicates the total number of packets lost during transmission from $a_i$ to $a_j$ within the same task. $P_{n}^{\text{rec}}$ is the total number of received packets by $a_j$ from $a_i$ in the $n$-th task, while $P_{n}^{\text{tra}}$ represents the number of those packets successfully forwarded by $a_j$. 




For EC devices, the outcomes of the tasks they execute are collected to measure their performance. An interaction $d_{a_i,b_m}$ captures the result of $b_m$ processing a task generated by $a_i$, where a successful execution is denoted by 1 and a failure by 0. Accordingly, an edge $e_{a_i, b_m}$ can be established from $a_i$ to $b_m$. The weight of $e_{a_i, b_m}$ represents the direct trust $T^{\text{dit}}_{a_i, b_m}$ that $a_i$ places in $b_m$, which is calculated as
\begin{align}
 \label{a_i_dir_b_m}
    T^{\text{dir}}_{a_i, b_m} = \frac{\sum^{N_{a_i,b_m}^{\text{tot}}}d_{a_i,b_m}}{N^{\text{tot}}_{a_i, b_m}},
\end{align}
where $N^{\text{tot}}_{a_i, b_m}$ is the total number of tasks executed by $b_m$ from $a_i$. From the historical interaction data, we ultimately construct an interaction graph that represents the direct trust relationships among devices, denoted as $G^{\text{dir}} = (\{{A}, {B}\}, {E}^{\text{dir}}, {W}^{\text{dir}})$ where ${E}^{\text{dir}}$ is the set of all edges, and ${W}^{\text{dir}} = \{\dots T^{\text{dir}}_{a_i,a_j}, T^{\text{dir}}_{a_i,b_m}\dots \}$ is the set of edge weights.

\textit{2) Propagation and Aggregation of Trust Information}:
To accurately evaluate the trustworthiness of devices within the system, trust information needs to be propagated and aggregated. The propagation process transfers trust values from one device to those that interact indirectly, thereby reflecting the mutual influence among devices. Through aggregation, trust data from various devices and interactions is consolidated to derive an accurate trust evaluation for each device. Therefore, we begin by generating initial embeddings for devices in $G^{\text{dir}}$ using node2vec\cite{A. Grover}, which maps each device $a_i$ or $b_m$ into a $D_a$-dimensional vector space, denoted as $h_{a_i}$ or $h_{b_m}$. Subsequently, device embeddings are learned from a global perspective by propagating and aggregating trust information.

Given that TF devices participate in task workflows by both receiving and forwarding tasks, they inherently function as trustors and trustees. Accordingly, their embeddings should effectively reflect this dual functionality. Specifically, in $G^{\text{dir}}$, a TF device’s out-degree represents the trust interactions it initiates (acting as a trustor), while its in-degree represents the trust interactions it receives (acting as a trustee). In addition, the trust values of devices need to be propagated to their $l$-hop neighbours, $l =1,\dots, L$. Therefore, we stack $L$ GNN-enabled propagation and aggregation layers, allowing each device to aggregate trust features from its neighbours. 


\textbf{{Trust propagation and aggregation of TF devices}}: 
When a TF device $a_j$ acts as a trustee, its $l$-order in-degree neighbours propagate their trust evaluation values to $a_j$. This process can be represented as follows
\vspace{-0.08 in}
\begin{align}
    \label{w_a_j_get_a_i}
    \omega_{a_j \gets a_i}^{(l)} &= W_{a_j \gets a_i}^{(l)} \cdot \chi_{a_j \gets a_i}^{(l)}, \\
    \label{mu_a_j_get_a_i}
    \mu_{a_j \gets a_i}^{(l)}  &= h_{a_i}^{(l)} \otimes \omega_{a_j \gets a_i}^{(l)},
\end{align}
where $ \chi_{a_j \gets a_i}^{(l)} \in \mathbb{R}^{D_T \times 1}$ is the embedding of trust value $T^{\text{dir}}_{a_i, a_j}$ that $a_i$ places in $a_j$, which is transformed by using binary encoding. $W_{a_j \gets a_i}^{(l)} \in \mathbb{R}^{D_a \times D_T}$ is a trainable transformation matrix, and $\otimes$ is a concatenation operation. Since $\mu_{a_j \gets a_i}^{(l)}$ encompasses the embeddings of trust and $a_i$, it can be interpreted as $a_i$'s recommendation for $a_j$. After receiving the messages from its $l$-hop neighbours, $a_j$ aggregates them. Then, we calculate the importance of neighbours to $a_j$ in order to assign varying weights to each neighbour. Furthermore, the weights are adjusted according to the frequency of interactions, as devices that interact more frequently with $a_j$ are considered to have higher importance.
The importance of each in-degree neighbour of $a_j$ is computed using device embeddings through an attention layer
\vspace{-0.06 in}
\begin{align}
    \bar{\psi}_{a_j\gets a_i} = \text{attention} (W^{(l)}_{a_j}h^{(l)}_{a_j}, W^{(l)}_{a_i} h^{(l)}_{a_i}),
\end{align}
where $W^{(l)}_{a_j} = W^{(l)}_{a_i} \in \mathbb{R}^{D_a \times D_a}$ are the shared linear transformation weight matrix. Then, we normalize $ \psi_{a_j\gets a_i}$ using the \textit{softmax} function
\vspace{-0.1 in}
\begin{align}
    \widehat{\psi}_{a_j\gets a_i} = \frac{\text{exp}(\bar{\psi}_{a_j\gets a_i})}{\sum_{a_i \in {\mathcal{N}}^{\text{in}}_{a_j}}\text{exp}(\bar{\psi}_{a_j\gets a_i})},
\end{align}
where ${\mathcal{N}}^{\text{in}}_{a_j}$ is the set of in-degree neighbors of $a_j$. Furthermore, the importance of each neighbour is weighted based on the number of tasks it assigns to $a_j$
\vspace{-0.1 in}
\begin{align}
  \widetilde{\psi}_{a_j\gets a_i} = \widehat{\psi}_{a_j\gets a_i} \frac{N_{a_i, a_j}}{\sum_{a_i \in {\mathcal{N}}^{\text{in}}_{a_j}} N_{a_i, a_j}},
\end{align}
where $\sum_{a_i \in {\mathcal{N}}^{\text{in}}_{a_j}} N_{a_i, a_j}$ represents the total number of tasks received by $a_j$ from its in-degree neighbors. The \textit{softmax} function is applied to normalize the weighted importance
\begin{align}
    \label{neighbor_weight}
   {\psi}_{a_j\gets a_i} = \frac{\text{exp}(\widetilde{\psi}_{a_j\gets a_i})}{\sum_{a_i \in {\mathcal{N}}^{\text{in}}_{a_j}} \text{exp}(\widetilde{\psi}_{a_j\gets a_i})}.
\end{align}
Leveraging the weighted importance, $a_j$ aggregates the messages from its in-degree neighbours to generate its embedding as a trustee, which is computed as
\begin{align}
    \label{h_a_j_te}
    h_{a_j}^{\text{te}(l)} = \sum_{a_i \in {\mathcal{N}}^{\text{in}}_{a_j}} {\psi}_{a_j\gets a_i} \mu_{a_j \gets a_i}^{(l)}.
\end{align}
When $a_j$ acts as a trustor, its embedding is obtained through a process similar to that when it acts as a trustee
\begin{align}
    \label{omega_a_j_to_a_i}
    \omega_{a_j \to a_i}^{(l)} &= W_{a_j \to a_i}^{(l)} \cdot \chi_{a_j \to a_i}^{(l)}, \\
    \label{mu_a_j_to_a_i}
    \mu_{a_j \to a_i}^{(l)}  &= h_{a_i}^{(l)} \otimes \omega_{a_j \to a_i}^{(l)},\\
    \label{h_a_j_tr}
    h_{a_j}^{\text{tr}(l)} &= \sum_{a_i \in {\mathcal{N}}^{\text{out}}_{a_j}} {\psi}_{a_j\to a_i} \mu_{a_j \to a_i}^{(l)},
\end{align}
where $W_{a_j \to a_i}^{(l)} \in \mathbb{R}^{D_a \times D_T}$ is a learnable transformation matrix, $\chi_{a_j \to a_i}^{(l)}$ is the embedding of direct trust value $T^{\text{dir}}_{a_j,a_i}$ that $a_j$ places in $a_i$, and ${\mathcal{N}}^{\text{out}}_{a_j}$ is the set of out-degree neighbors of $a_j$. 
To learn a more comprehensive embedding for $a_j$, the embeddings from its roles as both a trustor and a trustee are merged. This fusion enables the capture of the full trust dynamics of $a_j$, encompassing both its outgoing and incoming trust relationships. The merging process is performed via a fully-connected layer
\vspace{-0.045 in}
\begin{align}
   \label{h_a_j}
    h_{a_j}^{(l)} = \sigma \left( W_{h^{\text{te}}h^{\text{tr}}}^{(l)} \cdot  \left(h_{a_j}^{\text{te}(l)} \otimes h_{a_j}^{\text{tr}(l)} \right)  + b_{h^{\text{te}}h^{\text{tr}}}^{(l)} \right),
\end{align}
where $ W_{h^{\text{te}}h^{\text{tr}}}^{(l)}$ and  $b_{h^{\text{te}}h^{\text{tr}}}^{(l)}$ are the learnable parameters, $\sigma$ is a non-linear activation function. $h_{a_j}^{(l)}$ is the final embedding of $a_j$ after propagating and aggregating trust information from its $l$-order neighbors. 

\vspace{-0.05 in}
\textbf{{Trust propagation and aggregation of EC devices}}:
Since EC devices only process tasks from TF devices, they serve exclusively as trustees. The computational procedure, similar to that of TF devices acting as trustees, is given by
\begin{align}
    \label{nc_devices_35}
     \omega_{b_m \gets a_i}^{(l)} &= W_{b_m \gets a_i}^{(l)} \cdot \chi_{b_m \gets a_i}^{(l)}, \\
     \label{mu_b_m_get_a_i}
    \mu_{b_m \gets a_i}^{(l)}  &= h_{a_i}^{(l)} \otimes \omega_{b_m \gets a_i}^{(l)}, \\
    \label{nc_devices_37}
    h_{b_m}^{(l)} &= \sum_{a_i \in {\mathcal{N}}^{\text{in}}_{b_m}} {\psi}_{b_m\gets a_i} \mu_{b_m \gets a_i}^{(l)},
\end{align}
where $ W_{b_m \gets a_i}^{(l)}$ is a learnable matrix, and $\mathcal{N}^{\text{in}}_{b_m}$ is the set of in-degree neighbours of $b_m$. $h_{b_m}^{(l)}$ is the final embedding of $b_m$, obtained after propagating and aggregating trust information from its $l$-order neighbors. Finally, the propagation and aggregation layer outputs the final embeddings for each TF device and each EC device, represented as $h_{a_j}^{(L)}$ and $ h_{b_m}^{(L)}$, respectively. These embeddings capture local topological information and fuse trust information from $L$-hop neighbours. Specifically, the embedding $h_{a_j}^{(L)}$ of each TF device incorporates the fusion of its roles as both a trustor and a trustee.

\vspace{-0.06 in}
\textit{3) Trust calculation}:
 With the above trust propagation and aggregation, interaction-based direct trust information is encoded into the device embeddings. A multi-layer perceptron (MLP) is employed as the prediction model to estimate the trust value between any pair of devices
\vspace{-0.08 in}
 \begin{align}
    \label{h_a_i_rightarrow_a_j}
    h_{a_i \Rightarrow a_j}  &= \sigma \left(\text{MLP} \left( h^{(L)}_{a_i} \otimes h^{(L)}_{a_j}  \right) \right).
 \end{align}
The output of this step is the probability values. The trust of $a_i$ toward $a_j$ is the maximum value in $h_{a_i \Rightarrow a_j}$, calculated as $T^{\text{his}}_{a_i,a_j} = \max(h_{a_i \Rightarrow a_j})$. Similarly, we can obtain the trust value of $a_i$ toward $b_m$, $T^{\text{his}}_{a_i,b_m}$.

\vspace{-0.03 in}
\textit{4) Model training}:
To train the GNN model, a cross-entropy loss function is used to measure the difference between the predicted trust values and the ground-truth trust values. The objective function is formulated as
\vspace{-0.08 in}
\begin{align}
    \mathcal{L} = - \frac{1}{|{W}^{\text{trust}}|} \sum_{|{W}^{\text{trust}}|} \log \left(h_{a_i \Rightarrow a_j, T^{\text{dir}}_{a_i,a_j}}\right) + \lambda \parallel \Theta \parallel_2^2,
\end{align}
where $\Theta$ denotes all trainable model parameters, and $\lambda$ controls the $L_2$ regularization strength to prevent over-fitting.

\subsection{Task-Specific Resource Trust Evaluation}
After obtaining the historical trustworthiness of devices, it is necessary to evaluate their task-specific resource trustworthiness, i.e., $T^{\text{res}}_{a_i,a_j}$ and $T^{\text{res}}_{a_i,b_m}$. For TF devices, their resources trustworthiness is determined by the following three conditions: i) Idle status: $a_j$ must be idle, meaning it is not engaged in forwarding other tasks at the time; ii) Available storage: $a_j$'s available storage must be sufficient to temporarily store task $C$ before forwarding it to the next device; iii) Sufficient energy: $a_j$ must have enough energy to both receive and forward task $C$. 
The task-specific resource trustworthiness of $a_j$ for task $C$ is calculated as $T^{\text{res}}_{a_i,a_j} (C) = T^{\text{idle}}_{a_i,a_j} T^{\text{sto}}_{a_i,a_j}T^{\text{eng}}_{a_i,a_j}$,
where $T^{\text{idle}}_{a_i,a_j}$ is used to determine whether $a_j$ meets condition 1, and $T^{\text{sto}}_{a_i,a_j}$ is applied to assess whether $a_j$ meets condition 2. They are defined as follows
\vspace{-0.03 in}
\begin{align}
    T^{\text{idle}}_{a_i,a_j} = \begin{cases}
    1, &\text{idle}, \\
    0, &\text{otherwise}, \\
   \end{cases} 
    T^{\text{sto}}_{a_i,a_j} = \begin{cases}
    1, & a^{\text{sto}}_{j} \geq c^{\text{size}}, \\
    0, & \text{otherwise}. \\
   \end{cases}
\end{align}
where $a_j^{\text{sto}}$ is the available storage of $a_j$. $T^{\text{eng}}_{a_i,a_j}$ is employed to evaluate whether $a_j$ satisfies condition 3, which is given by
\vspace{-0.084 in}
\begin{align}
    T^{\text{eng}}_{a_i,a_j} = \begin{cases}
    1, & a^{\text{eng}}_{j} \geq E_{a_j}^{\text{rec}} + E^{\text{tra}}_{a_j} , \\
    0, & \text{otherwise}, \\
   \end{cases}
\end{align}
where $a^{\text{eng}}_{j}$ is the available energy of $a_j$. $E_{a_j}^{\text{rec}}$ and $E^{\text{tra}}_{a_j}$ represent the energy consumption of $a_j$ for receiving and transmitting task $C$, respectively. They are computed based on a first-order radio model. The task-specific resource trustworthiness of $b_m$ for task $C$ is calculated as $T^{\text{res}}_{a_i,b_m} (C) = T^{\text{idle}}_{a_i,b_m} T^{\text{sto}}_{a_i,b_m}T^{\text{eng}}_{a_i,b_m}$,
where $T^{\text{idle}}_{a_i,b_m}$ and $T^{\text{sto}}_{a_i,b_m}$ follow the same evaluation logic as $T^{\text{idle}}_{a_i,a_j}$ and $T^{\text{sto}}_{a_i,a_j}$. The calculation for $T^{\text{eng}}_{a_i,b_m}$ is given as
\begin{align}
    T^{\text{eng}}_{a_i,b_m} = \begin{cases}
    1, & b^{\text{eng}}_{m} \geq \epsilon (b_{m}^{\text{cpu}})^{2}c^{\text{des}}c^{\text{size}}, \\
    0, & \text{otherwise}, \\
   \end{cases}
\end{align}
where $b_m^{\text{eng}}$ is the available energy of $b_m$, $\epsilon (b_{m}^{\text{cpu}})^{2}c^{\text{des}}c^{\text{size}}$ is the energy consumption of $b_m$ when executing task $C$, $b_m^{\text{cpu}}$ is the CPU frequency of $b_m$, and $\epsilon$ is set to $10^{-11}$. 

\begin{algorithm}[h!]
        \footnotesize
	\caption{Path planning via A* search}
	\label{multi-hop path}
		\KwIn {$G^{\text{trust}}$, task initiator $a_i$}
            \KwOut{$\pi^{\text{max}}$}
             Initialize each $Ag_{b_m}$ and prepare a priority queue $Q_{b_m}$ for each $Ag_{b_m}$\\
             $Q_{b_m}.\text{push}(b_m, \text{path}=[b_m], f = T_{a_i,b_m})$\\
             Each $Ag_{b_m}$ performs the following search in parallel:\\
             \While{$Q_{b_m}$ \textnormal{is not empty}}{
                $\text{curr}, \text{path} \gets Q_{b_m}.\text{pop}()$ // Pop up the path with the largest $f$ value \\
                \If{\textnormal{curr} == $a_i$}{
                 Store path and calculate the average trust value \\
                 End search for this agent
                 }
                 \For{\textnormal{each neighbor $a_j$ in neighbors (curr)}}{
                       \If{$a_j \in$ \textnormal{path}}{
                            Continue
                       }
                       new\_path $\gets$ path + $a_j$\\
                       $f(a_j) = f_1(a_j) + f_2(a_j)$\\
                       $Q_{b_m}.\text{push}(a_j, \text{new\_path}, f(a_j))$
                 }
             }  
   $a_i$ selects the path $\pi^{\text{max}}$ with the highest average trust value from all the paths from EC devices.
\end{algorithm}

\vspace{-0.1 in}
\subsection{Multi-Hop Path Planning}

After obtaining the historical trustworthiness and the task-specific resource trustworthiness, equations (\ref{final_a_i_a_j}) and (\ref{final_a_i_b_m}) are applied to compute the task forwarding trust for all TF devices and the task computing trust for all EC devices, respectively. 
Subsequently, based on the trust thresholds defined by task $C$, all devices in the network topology $G^{\text{top}}$ that fail to meet the thresholds are excluded, resulting in a refined network topology $G^{\text{trust}}$ consisting solely of trusted devices. To identify the path with the highest average trust value from $a_i$ to one of EC devices in $G^{\text{trust}}$, an agent $Ag_{b_m}$ is deployed at each EC device $b_m$. Each agent executes the A* search algorithm to find the path from itself to $a_i$ that maximizes the average trust value. Assuming that one of the agents accesses $a_j$, the heuristic function of the A* algorithm is designed as
\begin{align}
    f(a_j) = f_1(a_j) + f_2(a_j),
\end{align}
where $f_1(a_j)$ represents the average trust value of the devices already visited along the path from an EC device $b_m$ to $a_j$, while $f_2(a_j)$ denotes the estimated trust value of the devices on the possible path from $a_j$ to task initiator $a_i$. $f_2(a_j)$ is calculated as the average trust value of the neighbor devices of $a_j$.
Eventually, $a_i$ receives $M$ paths from EC devices and selects the one with the maximum average trust value, denoted as $\pi^{\text{max}}$. The detailed algorithm is provided in Algorithm~\ref{multi-hop path}.

\section{Experimental Results}

\subsection{Experimental Setup}

In this study, we consider a face recognition task. The task involves processing a series of photographs, where an EC device is required to count the number of faces in the images.
The task size is 50 MB, with a processing density of 2,339 cycles/bit~\cite{J. Kwak}. The trust demands $c^{\text{TF}}$ and $c^{\text{EC}}$ are set to 0.4 and 0.3, respectively. Three types of devices are considered: two terminal devices (iPhone 15 and Pixel 8) and one EC device (Dell Edge 5200). Performance data related to task forwarding and computation is collected from these devices to build accurate device models. Then, we deploy 25 iPhone models, 25 Pixel 8 models, and 10 Dell Edge 5200 models using NS3 to simulate the collaborative system. A total of 5,000 tasks are executed, and the interaction data between devices is recorded to generate the dataset. In each task, a terminal device is randomly selected as the initiator, and the task is relayed via multiple hops to an EC device. $\alpha_1$ and $\alpha_2$ are set to 0.6 and 0.4, respectively. 
The initialized embeddings of devices are set to a dimension of 128. In terms of hyperparameters of GNN, $L = 3$ propagation and aggregation layers are used, with output dimensions of 32, 64, and 32 for the first, second, and third layers, respectively~\cite{W. Lin}. The learning rate is chosen from $\{10^{-4}, 10^{-3}, 5 \times 10^{-3}, 10^{-2}, 5 \times 10^{-2} \}$, and the coefficient of $L_2$ regularization is $10^{-5}$. The dropout rate is in $\{0, 0.1, 0.3, 0.5, 0.8 \}$. Xavier initializer is used to initialize the parameters of GNN. 
Following~\cite{G. Liu}, we split the dataset as $80\%$ and $20\%$ for training and testing sets, respectively. 
The GNN model is trained on an NVIDIA P100 GPU using the Google Cloud Platform. 

\begin{figure}[t!]
\centering
\includegraphics[scale=0.52]{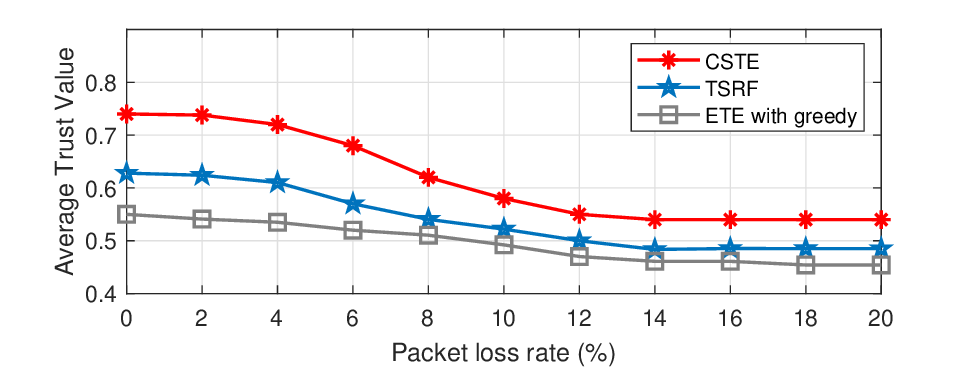}
\caption{Comparison of average trust value under different packet loss rates.}
\label{plr}
\end{figure}
\subsection{Impact of Packet Loss Rate}
We set the packet loss rate of two-thirds of the terminal devices to be the same and vary this packet loss rate. The average trust values of the obtained paths are compared with those produced by TSRF~\cite{J. Duan} and ETE~\cite{Y. Yu} with greedy. As shown in Fig.~\ref{plr}, when the packet loss rate is below 4\%, the average trust values of the paths obtained by the three methods remain relatively stable, because there are enough trusted terminal devices that can be selected in the system. When the packet loss rate exceeds 4\%, the average trust value of the paths obtained by CSTE decreases rapidly. This is because the number of terminal devices satisfying the trust demand decreases rapidly, resulting in a significant reduction in the number of devices that can be selected. When the packet loss rate exceeds 12\%, the average trust values of the paths obtained by the three methods stabilize again, as devices that do not meet the trust demand have been excluded, and the trust values of the remaining one-third of terminal devices are unaffected by changes in packet loss rate. In addition, CSTE consistently achieves the highest average trust value at different packet loss rates. In contrast, the approach that combines ETE with the greedy strategy yields the least favorable results. This is likely due to the fact that ETE focuses solely on trust evaluation, while the path selection relies on the greedy strategy, which is unable to identify globally optimal paths.

\vspace{-0.03 in}
\subsection{Impact of Task Forwarding Success Rate }

\begin{figure}[t!]
\centering
\includegraphics[scale=0.51]{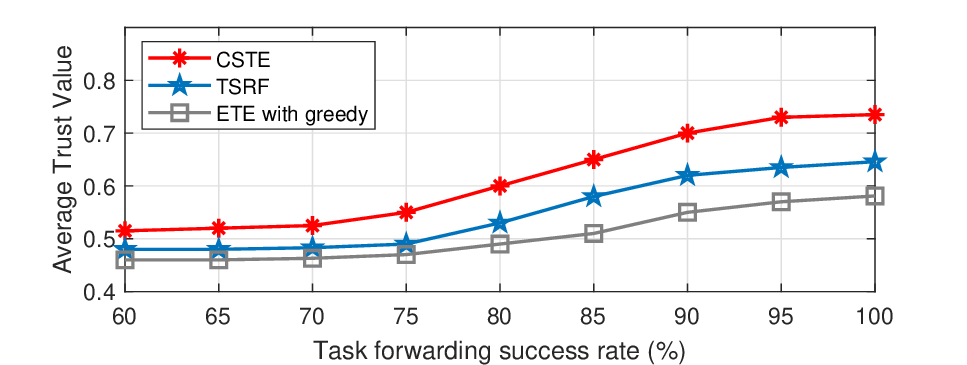}
\caption{Comparison of average trust value under different task forwarding success rates.}
\label{tfsr}
\end{figure}

Similar to the previous subsection, in this subsection we vary
the task forwarding success rate and observe the changes
in the experimental results. As shown in Fig.~\ref{tfsr}, when the task forwarding success rate is lower than 70\%, the average trust values of the paths obtained by the three methods remain almost constant. This is due to the fact that the terminal devices that do not satisfy the trust demand of task $C$ have been effectively identified, while the trust values of the remaining devices are not affected by the change in the task forwarding success rate. As the task forwarding success rate exceeds 70\%, the average trust value of the paths generated by CSTE increases rapidly. This improvement is attributed to the increasing number of terminal devices that satisfy the trust threshold, which allows more high-trust devices to be incorporated into the planned collaboration paths. Compared with two baseline algorithms, the proposed CSTE algorithm consistently yields paths with the highest average trust values across different levels of task forwarding success rate, demonstrating superior performance.

\vspace{-0.05 in}
\section{Conclusion}

This study investigates the problem of planning a trust-maximizing multi-hop cooperative path in complex systems. To solve this problem, the novel CSTE mechanism is proposed. The mechanism begins by utilizing a GNN-enabled trust evaluation framework to compute the historical trustworthiness of devices from stable historical interaction data. 
Next, a task-specific trust evaluation method is applied to assess the dynamic resource trustworthiness of devices based on task requirements. These two evaluation results are then integrated to identify trustworthy collaborators. Finally, the A* search algorithm is employed to efficiently plan a multi-hop collaboration path with the highest average trust. Extensive simulations demonstrate that CSTE consistently outperforms baseline algorithms by identifying multi-hop paths with the highest average trust values under various network conditions. 

\end{document}